# Spin current pumped by resonant skyrmion

Chunlei Zhang[1], Jianing Wang[1], Chendong Jin[1], Haiyan Xia[1], Jianbo Wang[1,2] and Qingfang Liu[1,*]

*[1]Key Laboratory for Magnetism and Magnetic Materials of the Ministry of Education, Lanzhou University, Lanzhou 730000, People's Republic of China.*

*[2]Key Laboratory for Special Function Materials and Structural Design of the Ministry of the Education, Lanzhou University, Lanzhou 730000, People's Republic of China.*

**Abstract**

Spin pumping is a widely recognized method to generate the spin current in the spintronics, which is acknowledged as a fundamentally dynamic process equivalent to the spin-transfer torque. In this work, we theoretically verify that the oscillating spin current can be pumped from the microwave-motivated breathing skyrmion. The skyrmion spin pumping can be excited by a relatively low frequency compared with the ferromagnetic resonance (FMR) and the current density is larger than the ordinary FMR spin pumping. Based on the skyrmion spin pumping, we build a high reading-speed racetrack memory model whose reading speed is an order of magnitude higher than the SOT (spin-orbit torque) /STT (spin-transfer torque) skyrmion racetrack. Our work explored the spin pumping phenomenon in the skyrmion, and it may contribute to the applications of the skyrmion-based device.

Keywords: Skyrmion, spin pumping, spin current, racetrack memory

___________________________

*Corresponding author: Qingfang Liu, liuqf@lzu.edu.cn

**Introduction**

Owing to the demands of spintronic research and applications, the exploration of methods for generating the spin current is an important issue attracting considerable attention. In magnetism, we usually utilize a microwave signal to resonate with the magnetic material (FMR), so the precessing magnetization induces a spin current at the FM (ferromagnet)-NM (normal metal) interface flowing into the NM[1,2]. Therefore, the current deriving from the momentum transfer of the excited spins (precessing magnetic moments) to the conduction electron via the exchange coupling is time-dependent, and then this mechanism existing in magnetic metallic multi-layers is the so-called spin pumping. Due to the technological potential of spin pumping, the researchers have made numerous experiments and calculations to complete the field, even in the non-linear spin texture like domain wall[3,4] and vortex[5].

The magnetic skyrmion is a kind of chiral topological protected spin texture and it attracts extensive interests since 2009 when it was found in the B20-type MnSi[6]. After that, the skyrmion is extensively found in other bulk material and multilayered ultrathin films[7]. Until now, it is still a novel topic to explore and acknowledged for the promising of the racetrack memory[8,9], oscillators[10,11], artificial neuron[12], logical gates[13] and so on. Due to these researches in recent years, we have already obtained a clear realization of the skyrmion dynamics, which is generally covered by the Thiele equation[14,15]. However, according to the Onsager's reciprocity relations[16], the momentum originating from the spin texture motion could also transfer to the conduction electrons and generate a spin current. So in this work, we explore the spin pumping phenomena in the microwave-motivated skyrmion. Through theory calculations and micromagnetic simulations, we show that the spin pumping of the skyrmion is influenced by the DMI and the microwave amplitude. Furthermore, based on these results, we devise a high reading-speed racetrack memory whose reading speed is much higher than STT or SOT induced skyrmion racetrack memory.

**Simulation model of the skyrmion spin pumping**

Figure 1(a) shows the simulation model, where the Neel skyrmion in the center of the upper layer resonates with the perpendicular microwave and injects spin current into the layer of heavy metal (HM). And the direction of microwave decides the skyrmion resonance mode—breathing mode. The micromagnetic simulations are performed by mumax3[17] and it numerically solves the LLG equation of magnetization dynamics

$$\frac{d\bm{m}}{dt} = -\gamma \bm{m} \times \bm{H}_{eff} + \alpha(\bm{m} \times \frac{d\bm{m}}{dt}) \quad (1)$$

where $\gamma$ is the gyromagnetic ratio, $\alpha$ is the Gilbert damping constant, and $\bm{H}_{eff}$ is the local effective magnetic field including the exchange field, anisotropy field, magnetostatic field, Zeeman field, and DMI field. The spectrum for skyrmion and uniformed FM layer in figure 1(b) used the customary variants: the exchange constant is adopted as $A = 15$ pJ/m; DMI constant is adopted as $D = 3.5$ mJ/m$^2$; $K = 800$ kJ/m$^3$ for perpendicular magnetic anisotropy; $M_s = 580$ kA/m as saturated magnetization and the coefficient $\alpha = 0.02$ for Gilbert damping (the dissipation from spin pumping is also contained in $\alpha$). The simulation size is $100 \times 100 \times 1$ nm$^3$ and divided into $100 \times 100 \times 1$ unit meshes, which is within the range of the Bloch exchange length[18]. As a result of the strong perpendicular magnetic anisotropy, the FMR of a uniformed FM layer hold a relatively high frequency (58.0 GHz) while the skyrmion resonant frequency (6.9 GHz) is far lower.

**Results and Discussion**

It is generally known that the spin pumping term is included by the Gilbert term and the spin current density is governed by the equation[2]

$$\begin{aligned} J_{sp}\bm{\sigma}(t) &= \frac{\hbar}{4\pi} A_r (\bm{m} \times \frac{d\bm{m}}{dt}); \\ J_{dc}\bm{\sigma}(t) &= \frac{1}{T} \int_T J_{sp}\bm{\sigma}(t) dt, \end{aligned} \quad (2)$$

where $\bm{m}$ is the unit vector along the local magnetization, $A_r$ is defined as the spin pumping conductance, $\bm{\sigma}(t)$ denotes the direction of the spin polarization, $J_{sp}$ is the amplitude of the spin current and the time average of $J_{sp}$, namely $J_{dc}$, is the DC component of the spin current. According to the symmetry of skyrmion, the magnetization of the skyrmion structure can be written as $\bm{m}(\bm{r}) = (cos\phi(\varphi)sin\theta(r), sin\phi(\varphi)sin\theta(r), cos\theta(r))$, and the in-plane spin current of the whole material is governed by the symmetry

$$\int_0^r \int_0^\pi \left(\mathbf{m} \times \frac{d\mathbf{m}}{dt}\right)_i dr d\theta = -\int_0^r \int_\pi^{2\pi} \left(\mathbf{m} \times \frac{d\mathbf{m}}{dt}\right)_i dr d\theta, \quad (i = x);$$

$$\int_0^r \int_{-\frac{\pi}{2}}^{\frac{\pi}{2}} \left(\mathbf{m} \times \frac{d\mathbf{m}}{dt}\right)_i dr d\theta = -\int_0^r \int_{\frac{\pi}{2}}^{\frac{3\pi}{2}} \left(\mathbf{m} \times \frac{d\mathbf{m}}{dt}\right)_i dr d\theta, \quad (i = y).$$

(3)

It is obviously that the spin current injected into the metal layer will diffuse in HM layer, so the net in-plane spin current density of the whole skyrmion is zero. We specially just calculate half of the plane as an estimation of the in-plane spin current density, so the spin density for skymion spin pumping can be calculated by the following equation

$$J_i \sigma_i(t) = \begin{cases} \int_0^r \int_0^\pi J_{sp}\sigma_i(t) dr d\theta \Big/ \xi, & (i = x); \\ \int_0^r \int_{-\frac{\pi}{2}}^{\frac{\pi}{2}} J_{sp}\sigma_i(t) dr d\theta \Big/ \xi, & (i = y); \\ \int_0^r \int_0^{2\pi} J_{sp}\sigma_i(t) dr d\theta \Big/ \xi, & (i = z); \end{cases}$$

$$J_{dc}\sigma(t) = \frac{1}{T} \int_T J_i \sigma_i(t) dt;$$

(4)

where the constant $\xi$ equals to the integral of whole surface. Furthermore, in this work, the DC component of the spin current we talked about only indicates to $i = z$.

Based on Eq. (4), we calculate the in-plane spin current density $J_x$ ($0 < \theta < \pi$) and out-of-plane spin current density $J_z$ as shown in Figure 2. The phase of $J_x$ in counter radial direction (figure 2(b)) has a difference of $\pi$, and this confirmed our theoretical speculate of Eq. (3). As shown in figure 2 (e) and (f) the resonant skyrmion can pump a spin current several times stronger than the FMR, while the spin current density of the uniformed FM state motivated at 6.9 GHz (skyrmion resonant frequency) is nearly zero. It is known that the $J_{sp}$ of Eq. (2) can be also written in

$$J_{sp}\sigma(t) = \frac{\hbar\omega}{4\pi} A_r \sin^2 \theta_M,$$

(5)

where $\omega$ is the microwave frequency and $\theta_M$ is the cone angle of precession. So the most intuitive causes of the enhancing of spin pumping is that the magnetization momentum of skyrmion breathing process is flipping while the ordinary FMR is just a small-angle precession, $\theta_M^{skyrmion} > \theta_M^{FMR}$. Respectively, the crests and troughs correspond to the maximum and minimum of skyrmion breathing radius. In figure 2(a) and (c), the spatial amplitude of spin current density, attained by Fourier transform, suggest that the skyrmion spin pumping is localized and limited in the skyrmion area.

It can be realized that the localized skyrmion spin pumping is independent from the external disturbance. Therefore, this observation may support a hypothesis that skyrmion lattice will help to enhance the spin pumping. We then calculate the current density of hexagonal skyrmion lattice (inset of figure 3) via the periodical boundary condition in the software mumax3. The result of figure 3 shows that the current density of skyrmion lattice can be substantially enhanced by an adjacent lattice constant *a*. However, the over small lattice constant will also suppress the skyrmion breathing and the current density decline rapidly when $a < 35$ nm, $D = 3.5$ mJ/m$^2$. For the skyrmion lattice with different DMI constants, the system has a different maximum of spin current, which suggests that the DMI should be taken into the research of skyrmion spin pumping.

As we all know, the DMI is one of the most important variants to form skyrmions in the magnetic systems and indispensable for a small skyrmion[19]. For the skyrmion, we usually define the helicity by the phase $\gamma$[20] appearing in

$$\phi(\varphi) = N_{sk}\varphi + \gamma,$$
$$N_{sk} = \frac{1}{4\pi} \iint d^2 \boldsymbol{r} \boldsymbol{m} \cdot \left( \frac{\partial \boldsymbol{m}}{\partial x} \times \frac{\partial \boldsymbol{m}}{\partial y} \right), \qquad (6)$$

where the $N_{sk}$ is defined as the skyrmion number. In the studies of the recent years, it is found that the skyrmion helicity can be decided by the type of DMI and the skyrmions with different helicities are called by a joint name—the twisted skyrmion. The hybrid DMI of twisted skyrmion considered in C4 symmetry can be written as[21,22]:

$$\boldsymbol{D}_{\text{Hybrid}} = \boldsymbol{D}_{\text{bulk}} \cos\Omega + \boldsymbol{D}_{\text{inter}} \sin\Omega,$$
$$\Omega = 90° - \gamma. \qquad (7)$$

Respectively, the $\boldsymbol{D}_{\text{bulk}}$ and $\boldsymbol{D}_{\text{inter}}$ represent the ordinary DMI interaction in the bulk and interface. The hybrid angle $\Omega$ denotes the angle between the DMI vector and the line of the two nearest spins. The twisted skyrmion will be a Neel(Bloch) skyrmion whose helicity is 0° (90°), when $\Omega$ equals to 90° (0°). Figure 4(a) shows that the out-of-plane spin current density reaches the maximum when the $\Omega$ comes to 90°. Although the change brings by $\Omega$ is slightly for the whole current $J_z$, the DC component $J_{dc}$ can be increased by 17.8% from $6.74 \times 10^{-9}$ to $8.20 \times 10^{-9}$ arb. unit. Actually, the phenomenon is caused by the changing of radius for different $\Omega$. As shown in the figure 4(b), the skyrmion radius is influenced by the increase of

Ω. As we mentioned in the beginning that the skyrmion spin pumping is related to the breathing scope, so we defined the difference δR between the maximum and minimum of skyrmion radius

$$\delta R = R_{max} - R_{min}. \tag{8}$$

The δR has a distinct advantage in describing the scope of skyrmion breathing process, and it rises as the Ω increasing. The dependence of δR and Ω indicates that the $J_{dc}$ is controlled by Ω through the δR. By the way, the definition of δR also assists us in the following research.

In comparison with the Ω of DMI, the DMI strength plays a more significant role in the skyrmion spin pumping. Due to the increase of δR shown in figure 5(a) noted in red, the in-plane pumped current $J_x$ also increases rapidly with the strength of interfacial DMI. We can know from the inset of figure 5(a) that the skyrmion radius is gradually increasing with the increase of DMI strength and limitation of the material size is non-negligible. Comparing the magnetization components at the *x*-axis (inset of figure 5(b)) for $D = 3.3$ mJ/m$^2$, the in-plane magnetization $m_y$ is apparently restricted by the square edge. Therefore, the increasing of out-of-plane current $J_z \propto f(m_x, m_y)$ and the DC component $J_{dc}$ tends to stop after a short growth (figure 5(b)). For the skyrmion in the confined area, the current density of different polarization directions shows different relations of the DMI constant.

As for the skyrmion, even the generally magnetic bubble in experiment, the field amplitude is also a common means to adjust the skyrmion radius, so the field amplitude plays an important role in the skyrmion breathing. And changing the amplitude of the microwave field is also the most convenient method for application to control the spin current density. So we calculate the skyrmion spin pumping current density for different microwave amplitude (figure 6(a)). Because the δR is linearly increasing with the microwave amplitude (inset of figure 6(a)), the $J_z$ and $J_x$ are also linearly related to the microwave amplitude *h*. The similar linear relation between $J_z$ ($J_x$) and *h* is also found in the common FMR spin pumping[23]. However, it has been verified that the DC component of the spin current in the ferromagnetic material can be described by the relation[23,24]:

$$J_{dc} \propto h^2. \tag{9}$$

The skyrmion spin pumping shows the same dependence with the microwave—the DC component of spin current $J_{dc}$ fits well with the quadratic curve in figure 6(b). These results denote that the method controlling the spin current by microwave can also be applied to the skyrmion spin pumping. And the skyrmion spin pumping with nice controllability shows a possibility for the spin current application.

As we all know, the skyrmion is promising for the racetrack memory, but the reading speed of racetrack memory is limited by the skyrmion velocity, driving by STT or SOT, $v \sim 100$ m/s. However, with the assistance of skyrmion spin pumping, we design a new multilayer skyrmion racetrack memory (figure 7(a)) with higher reading speed. For the layers from the top to the bottom, the model is respectively composed of the FM layer, HM layer where the isolator is at the center to prevent the in-plane current $J_x$ of different parts, ($0 < \theta < \pi$) and ($\pi < \theta < 2\pi$), from mixing, and the PZT layer for the propagation of acoustic wave. As the sectional view at the upper left corner of figure 7(a) shows, the spin current can be detected by the voltage $V_{ISHE}$ of inverse spin Hall effect and the relation between $V_{ISHE}$ and $\sigma$ is therefore given by:

$$V_{ISHE}^i = \rho l \left( \theta_{SH} \frac{2e}{\hbar} \boldsymbol{J}_i \times \boldsymbol{\sigma}_i \right), (i = x, y, z). \tag{10}$$

The $\boldsymbol{J}_i$ is denoted as the amplitude and flowing direction of the spin current, $u$ is the distance of the two nearest data restore in the racetrack, $l$ is the material length along electric field, the spin Hall angle $\theta_{SH}$ is set as 0.01 and $\rho = 2.22 \times 10^{-7}$ Ω·m is the resistivity of the metal layer. According to the Eq. 10, we take $i = x$, corresponding to the polarization direction $\boldsymbol{\sigma}_x$, to generate a voltage $V_{ISHE}^y$ in the $y$-direction. Owing to the problem that the net in-plane current density shown in figure 2(a) would be mixed to zero, an insulator layer is set in the center to separate the two parts calculated by Eq. (2) and (4). So the voltage in HM layer can be derived from the model and equation, $V_{ISHE}^y \sim 0.01$ μV.

As we mentioned above, the frequency motivating the skyrmion spin pumping cannot motivate the resonance of material in the ferromagnetic state, whose pumped current is almost zero. When a skyrmion racetrack memory is motivated by a short microwave field pulse of the skyrmion spin pumping frequency, the data point on the racetrack, denoted by blanks/skyrmions corresponding to 0/1, will cause a transient low/high voltage by the ISHE. The high voltage in the *y*-direction at data points can also cause a transient deformation of the PZT track shown as the $t_0$ of figure 7(b), the peaks and blanks in the PZT track denote data string 111101011011 and the data distance $u$ is set to 100 nm to avoid the disturbance from neighbor data points. We make an approximation that the deformation is uniformly applied in the scope of 10 nm on the edge and ignore the energy dissipation, the deformations will spread along the track (*x*-axis) in the acoustic wave velocity of 3018.8 m/s. After t = 533 ps, the deformation peaks are mixed, but the data are still contained in the wave. The peaks of data 1 are mixed to four wave packets and the three major troughs denote the former data 0. By setting a deformation threshold denoted by the gray solid line, the wave packets and troughs can be easily translated to the data string 111101011011. Because the deformation in the PZT can also be converted to a voltage, a sensitive voltage detector with an adjacent voltage threshold can be responsible for the data reading. Our design of the high reading-speed racetrack memory model takes advantage of the acoustic wave and spin pumping, which bypasses the limitation of skyrmion velocity and provides a new thought of the racetrack memory.

**Conclusions**

In conclusion, the purpose of the study is to investigate the skyrmion spin pumping phenomena and explore its application prospect. Firstly, the skyrmion spin pumping can be assessed by simulation and calculation. It shows that the skyrmion spin pumping hold a lower excitation frequency and the density of the pumped current is stronger than the ordinary spin pumping by FMR. Then, we discuss how to control the skyrmion spin pumping by adjusting Dzyaloshinskii-Moriya interaction (DMI) and the amplitude of the microwave, and it is found that: (i) the DC spin current density could be influenced by the hybrid angle $\Omega$ of the DMI, the current density of $\Omega = 90°$ have a growth of 17.8% than that of $\Omega = 0°$; (ii) the in-plane spin current density linearly increases as the strength of DMI increasing, while the out-of-plane spin current density is limited by the material size; (iii) DC spin current density rises with the microwave amplitude as a quadratic function, $J_{dc} \propto h^2$, while the amplitude of the mixed current grow with microwave linearly. At last, we build a high reading-speed racetrack memory model. The data record by skyrmions can be transformed to the deformation of the

PZT by the ISHE voltage and the deformation can spread in a velocity of 3018.8 m/s larger than the skyrmion velocity driven by STT or SOT. The work provides some of the new information about the researches of spin pumping and skyrmion, and it may help researchers to get some thoughts about the skyrmion-based devices.

**Acknowledgments**

This work is supported by National Science Fund of China (11574121 and 51771086).

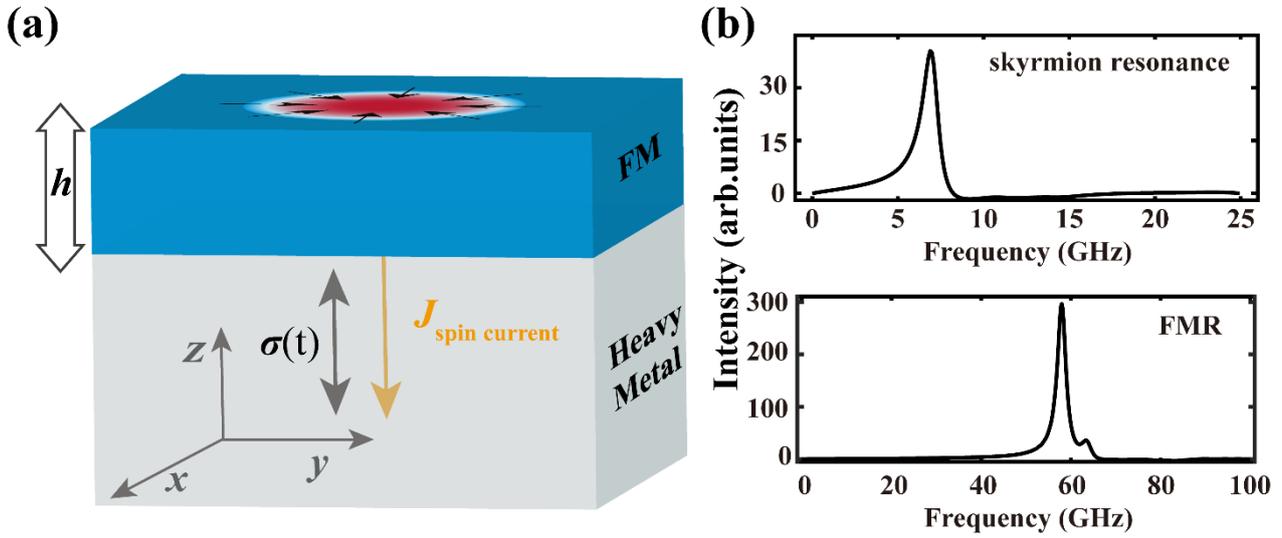

Fig. 1. (a) Schematic of the skyrmion spin pumping model. The Neel skyrmion in FM layer is resonant with the microwave perpendicular to the x-y plane and a current noted in yellow is pumped into HM. The solid black arrow denotes the time-dependent spin-polarization direction $\sigma(t)$. (b) Spectrum of skyrmion and uniformed FM layer for $D$ = 3.5 mJ/m$^2$, $h$ = 0.5 mT. Respectively, the resonant frequency is 6.9 GHz and 58.0 GHz for skyrmion and uniformed FM layer.

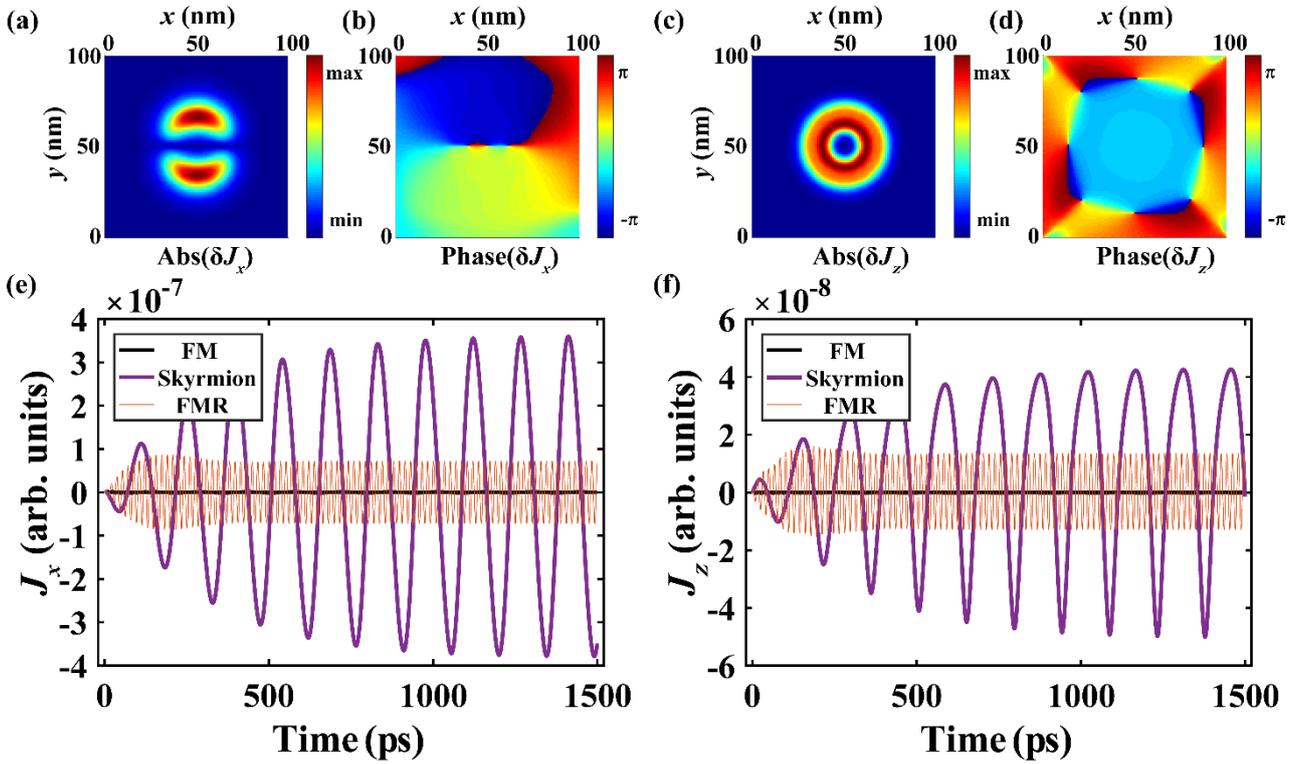

Fig. 2. Spatial maps of the skyrmion spin pumping current density in the polarization direction of $x$ ((a), (b)) and $z$ ((c), (d)). (e) and (f) shows the skyrmion spin pumping current density $J_x$ and $J_z$ at the resonant frequency of 6.9 GHz. The current pumped by skyrmion is denoted by the purple line, the orange fine line denotes spin pumping by FMR and the current of FM state excited in 6.9 GHz is denoted in black.

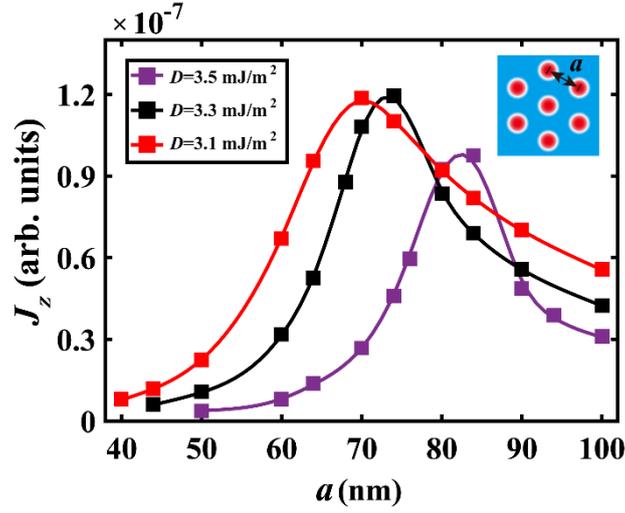

Fig. 3. The relation between the density of pumped current $J_z$ and the skyrmion lattice constant $a$ for different DMI constants.

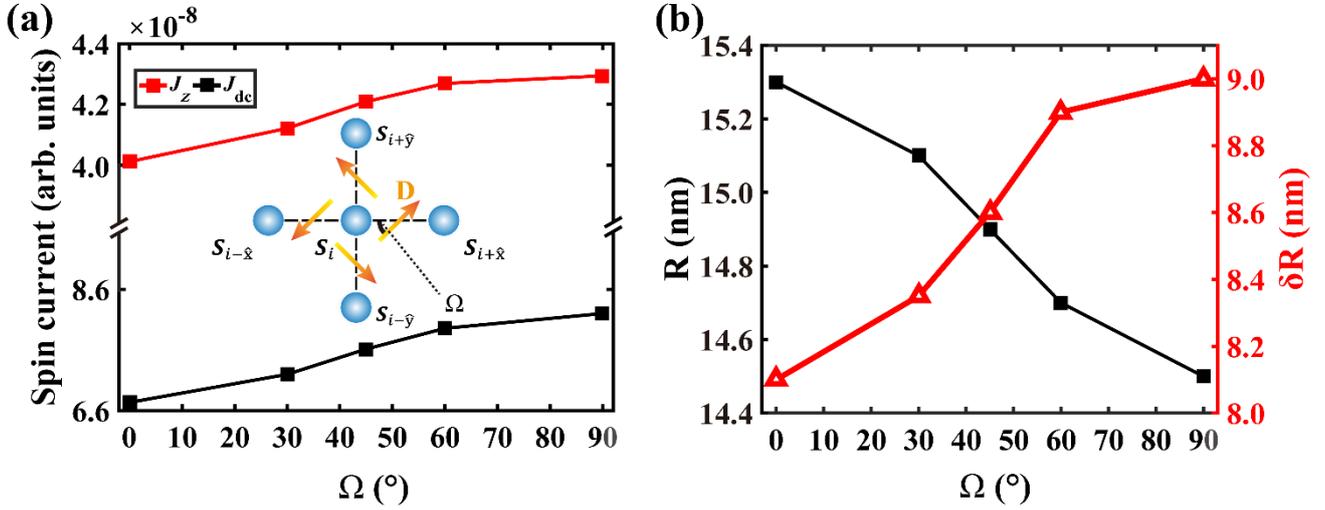

Fig. 4. (a) The dependence between the current density and the hybrid angle $\Omega$. (inset) Schematic of the hybrid angle $\Omega$ and the DMI vectors are denoted in orange arrows. (b) The skyrmion radius and the breathing radius as a function of the hybrid angle.

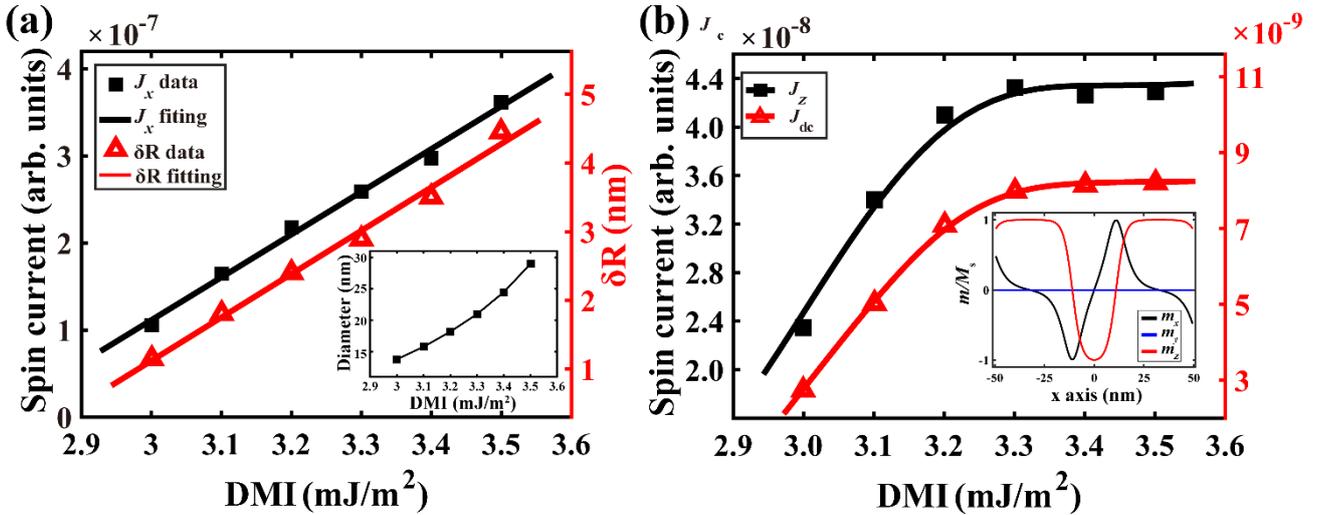

Fig. 5. (a) The relation of in-plane spin current density $J_x$ and interfacial DMI constant in a $100 \times 100 \times 1$ nm³ square. The inset shows the skyrmion diameter for DMI constant from 3.0 to 3.5 mJ/m². (b) The out-of-plane spin current density $J_z$ and $J_{dc}$ as a function of DMI

constant. (inset) The spatial profiles of the local magnetization along the *x*-axis.

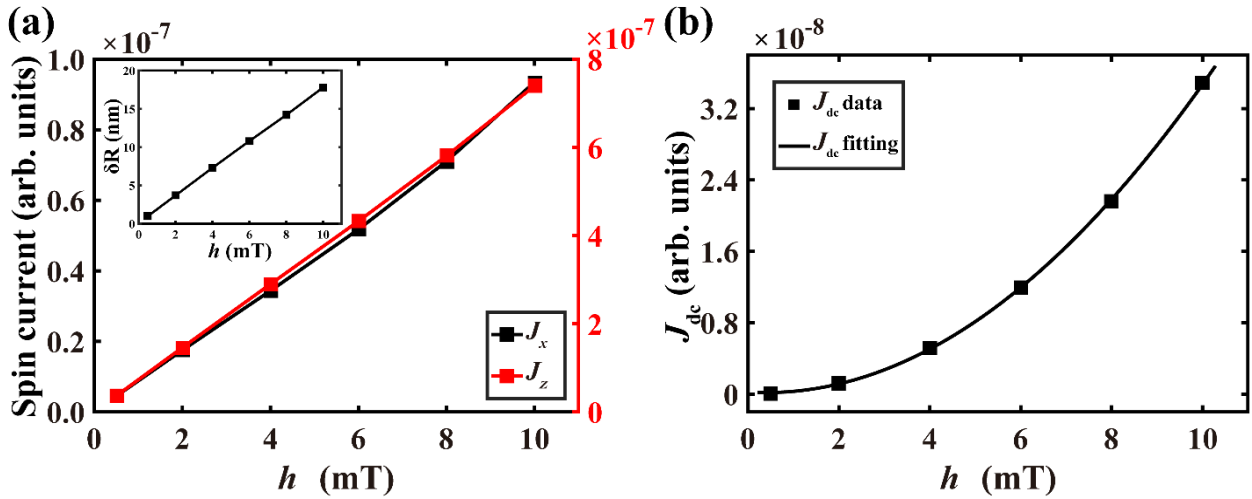

Fig. 6. (a) The relation between the out-of-plane/in-plane spin current density and the microwave amplitude. The inset displays the δR as a function of microwave amplitude. (b) The DC component of the current density as a function of microwave amplitude and the data point are fitted by a quadratic curve.

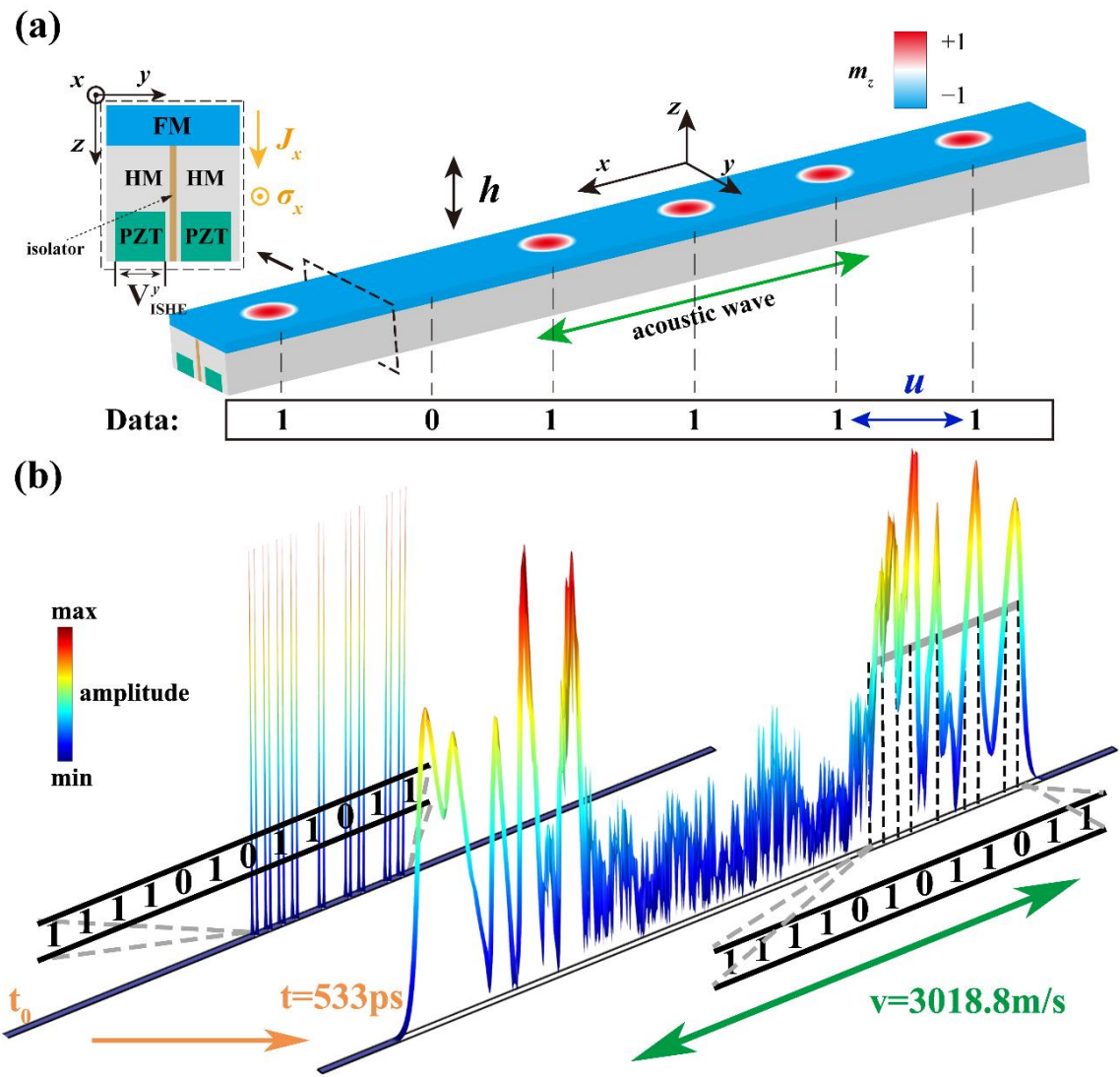

Fig. 7. (a) The model of the high reading speed structure built on the racetrack memory. The skyrmions and blanks denote data of 1 and

0, while the data gap is denoted by *u*. The brown color in the center of the HM layer shown in the three-dimensional model is the isolator. The inset map in the top left is the *y-z* plane map. The $\mathbf{J}_x$ is divided into two parts, $\mathbf{J}_x$ (*y* > 0) and $\mathbf{J}_x$ (*y* < 0), by the isolator, so the spin current $\mathbf{J}_x$ (*y* > 0)/$\mathbf{J}_x$ (*y* < 0) will lead to the voltage $V_{ISHE}^y$ applied on the PZT layer noted in green. (b) A microwave pulse is applied to the model. For $t_0$ of figure (b), the skyrmion spin pumping voltage causes deformation peaks at the PZT track and the 9 peaks and 3 blanks denote the data string 111101011011. The color bar shows the deformation amplitude of the PZT track. After t = 533ps, it is shown in the right that the deformations spread as an acoustic wave in *v* = 3018.8 m/s. The gray solid line denotes the reading threshold. The black dot lines denote that the periodically read data point over the threshold and the string is still the 111101011011 as $t_0$.